\documentclass[twocolumn,showpacs,preprintnumbers,amsmath,amssymb]{revtex4-1}

\usepackage{graphicx}
\usepackage{dcolumn}
\usepackage{bm}

\begin{document}

\preprint{APS/123-QED}

\title{INFLUENCE OF DEFORMATION ON FRACTAL DIMENSION OF METALS STRUCTURE}

\author{ANATOLIY ZAVDOVEEV, YAN BEYGELZIMER, VICTOR VARYUKHIN, BORIS EFROS}

 \email{zavdoveev@fti.dn.ua}
\affiliation{Donetsk Institute of Physics and Engineering, Ukrainian
Academy of Sciences,
\\83114, R.Luxemburg str. 72, Donetsk, Ukraine
}%

\date{\today}

\begin{abstract}
The image fractal analysis is actively used in all
 science branches. In particular in materials science 
the fractal analysis is applied to study microstructure 
of deformed metals because its structure can be interpreted
 as the fractal image. It is well known that such images 
can be described by fractal dimension. In this paper, 
the fractal dimension change for different kinds of 
metals in the processes of severe plastic deformation 
(SPD) is explored. It is shown that for the
 undeformed metals the stochastic network of 
structural elements boundaries has the fractal 
dimension $D_{b}=1.60\pm0.03$. The SPD 
leads to increasing the fractal dimension 
up to $D=1.80\pm0.03$. Possible reasons of changes i
n fractal dimension of metals structures in the 
SPD processes are analyzed. 
\end{abstract}

\pacs{61.43.Hv; 81.40.-z}
\maketitle

\section{Introduction}

It is known that severe plastic deformation (SPD) 
allows to get materials with sub micro- and nanostructures 
\cite{via00, zl02}, possessing 
principally new complex of properties. However, physical principles 
of creating such materials haven’t been exposed completely. On this 
account it is actual to explore the evolution of materials structure 
in the SPD processes. In materials science, the fractal analysis and 
multifractal parameterization are actively used \cite{vkb01} as deformed metals structures represent a kind of a self-similar set with grains and grain-boundary generating a network. Such network can be interpreted as the fractal image of the structure. As is generally known (Mandelbrot, 1983; Feder, 1991), such image is characterized by fractal or the so-called Hausdorf dimension which is an important quantative description of the explored objects. 
 
The basic hypothesis of the given work is that evolution of metals
 structure occurs in a self-similar manner with the formation of
 fractal structures \cite{bvos03,b05},
 that substantially simplifies their description. This hypothesis
 was proved theoretically: assuming that a successive set of 
high-angular boundaries is prefractal of the same fractal, we 
have the following estimation for the area of the given set in 
unit volume of material: $S~d-n$, where $d$ - characteristic size 
of fragment, $n = D - 1$, and $D$ - fractal dimension of lines set 
on the plane crossing high-angular boundary. The $D$ value lies 
within the limits of $1<D<2$. If the sizes of prefractal elements 
are distributed in a wide range, the dimension of fractal 
substantially differs from $2$. If the sizes of elements are 
approximately identical, the dimension of fractal approaches $2$. 
According to \cite{b05} at the self-similar stage 
of fragmentation $n = const$, with value lying between $0$ and $1$,
 and at a final stage, when the sizes of fragments approach 
the minimum, $n 1$. So, the main purpose of this work is to 
examine above listed hypothesis experimentally
 
\begin{figure}
\hspace{0.06 cm}
\includegraphics [width=3 in] {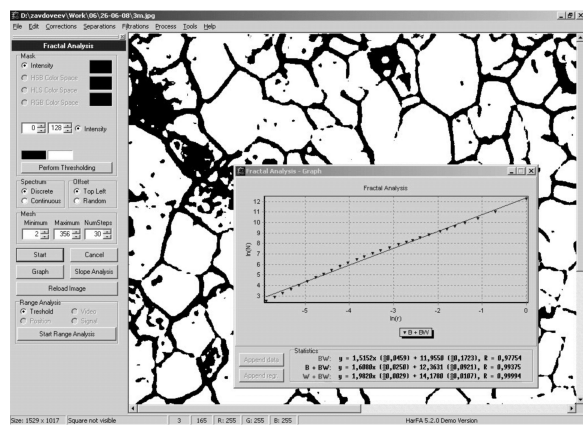}
\caption{\label{f1} Interface of HarFA 4.9 soft ware}
\end{figure}

\section{Materials and method}

\begin{figure}
\hspace{0.06 cm}
\includegraphics [width=3 in] {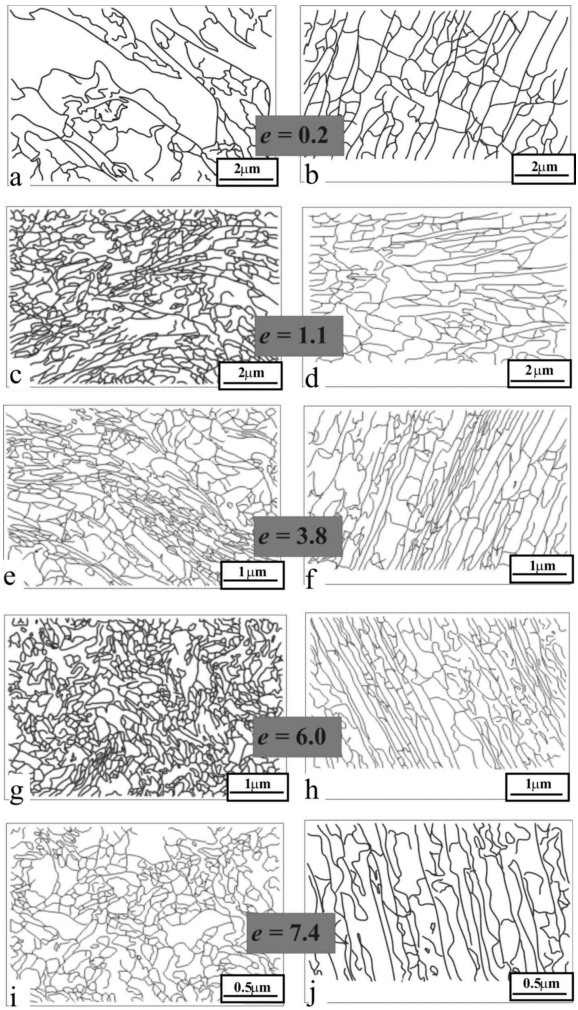}
\caption{\label{f2} Approximation of iron wire
microstructure drawn to different plastic strain
(Langford G. A, 1970). Cross section (a,c,e,g,i)
and longitudinal section (b,d,f,h,j)}
\end{figure}

\begin{figure}
\hspace{0.06 cm}
\includegraphics [width=3 in] {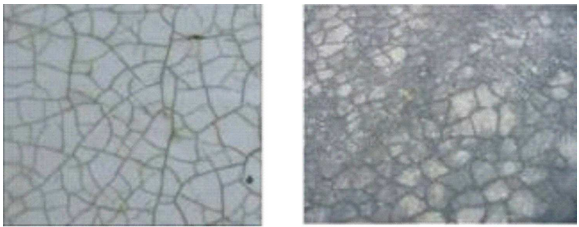}
\caption{\label{f4} Model nonmetal objects. Glaze (left),
asphalt (right)}
\end{figure}

The theory of self-similar sets 
\cite{bp01} is the basis 
of the fractal analysis method. They can be 
got by transformation of initial (half-tone) 
images of structures to their black and white (binary) ones. 
This procedure was carried out by the special programs Adobe 
Photoshop $7.0$, Corel Draw $11$, Image Tool $2.0$. In the given work the 
images were got by digitizing microstructures means of a scanner. 
There is a principle distinction in the nature of half-tone and binary 
pictures of structures. On microscope screen the half-tone picture 
is constructed by signals of variable amplitude. So, the contours 
of objects of the real material structure are determined by 
conditions of sample preparation (polishing and etching the specimen) 
and have the physical ground. On monitor the type of binary pictures 
is determined by the threshold of transformation and depends on 
the quantity of single pulses (zeros and unities) of constant 
amplitude. Thus, the binary image should be considered as a 
discrete approximation of characteristic functions of an object 
or as a statistically-geometric model of the structure. With 
the successful choice of the threshold of transformation the 
operation of binarization does not bring in substantial distortions 
in statistical data files and conserves scale correspondence in the 
geometrical parameters of the analyzed objects. Therefore with the 
visual coincidence of images it is considered that the results of 
measuring on the black and white images characterize the observed 
objects of the real structures \cite{c77}. The explored 
objects were the approximation of glaze and asphalt structures (fig.\ref{f4}); 
approximation of iron wire (fig.\ref{f2})
\begin{figure}
\hspace{0.06 cm}
\includegraphics [width=3 in] {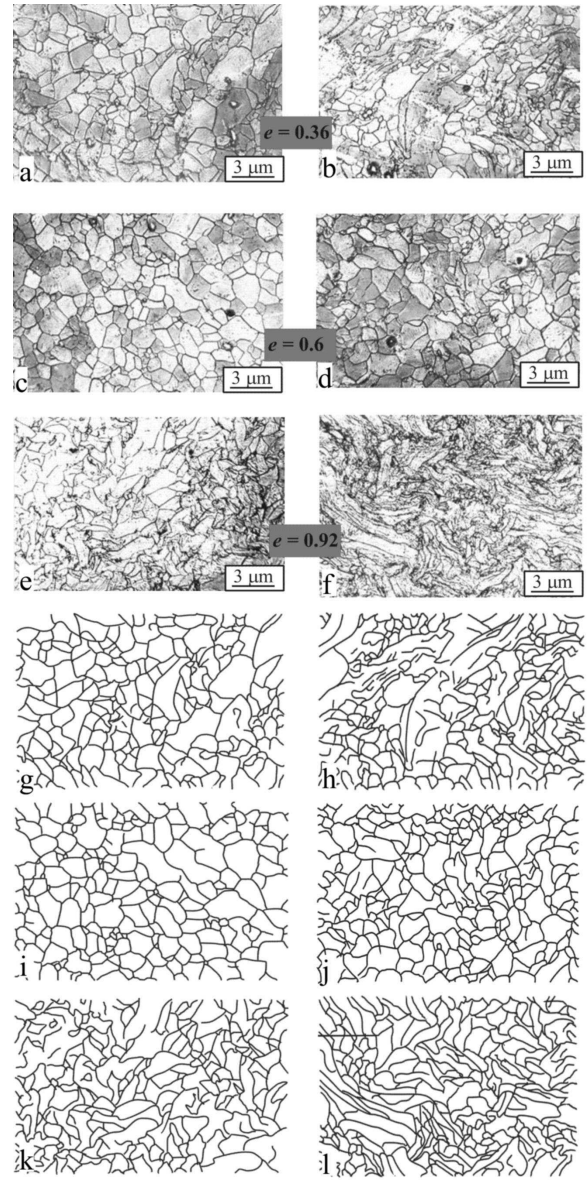}
\caption{\label{f3} Microstructure
of molybdenum deformed to different plastic
strain (Pbp=0.1 MPa a,c,e;Pbp=900 MPa b,d,f) and the approximation
(Pbp=0.1 MPa g,i,k;Pbp=900 MPa h,j,l)}
\end{figure}
 and molybdenum (fig.\ref{f3}) microstructures, 
deformed by different methods and to different stress \cite{e85,l70}. To
count the fractal dimension the so-called “Box counting method” was used. 
It works by covering fractal (its image) with boxes (squares) 
and then evaluating how many boxes are needed to cover fractal 
completely. Repeating this measurement with different sizes of 
boxes will result in logarithmic function of box size (x-axis) 
and number of boxes needed to cover fractal (y-axis). The slope 
of this function is referred as
box dimension. Box dimension is taken as an appropriate 
approximation of fractal dimension,   complete explanation 
is \cite{z01}. The HarFA $4.9$ program is in t
he basis of this method developed in the Institute of Physical 
and Applied Chemistry, Technical University of Brno in Czech Republic. 
The given software has been tested on regular fractals, for 
example the Serpinsky carpet $(D=1.79)$. More detailed description
 of the given program can be found in \cite{k05}. The 
general interface of the program is presented in fig.\ref{f1}. The 
got binary images of the explored metals structures were 
analyzed by the HarFA $4.9$ program. In the process of analysis 
the pattern area structure was varied from $50$ percent to $200$ percent
 with the step of $50$ percent.
 The pattern was also divided into four parts and the obtained constituents were processed

\section{RESULTS AND DISCUSSION}

In this paper were analyzed different kinds of 
materials both metals and nonmetals. As nonmetals there were 
model objects glaze and asphalt, they are presented in Fig.\ref{f4}. 
These are natural objects and generated by them network of fragments 
boundary is generated in stochastic manner. Also we analyzed low deformed 
(plastic strain $e<0.4$) metals: iron wire and molybdenum presented in 
Fig.\ref{f2} and Fig.\ref{f3} accordingly. 

\begin{figure}
\hspace{0.06 cm}
\includegraphics [width=3 in]{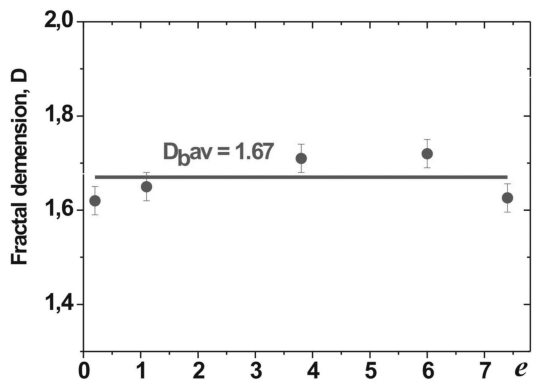}
\caption{\label{f5} Fractal dimension vs. plastic strain for
iron wire. Longitudinal section}
\end{figure}

\begin{figure}
\hspace{0.06 cm}
\includegraphics [width=3 in]{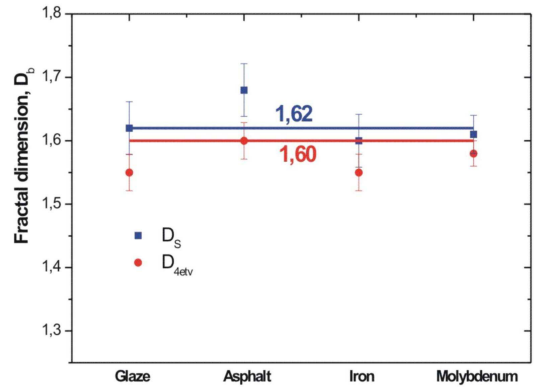}
\caption{\label{f6} Fractal dimension diagrams for different
kinds of material}
\end{figure}

We can state that for both the nonmetal natural objects and law-deformed 
metals the fractal dimension remains constant $D=1.60\pm0.03$ (Fig.\ref{f5}) 
This observation evidences that at low deformations or at recrystallization of 
metal the process of structure formation is probable. In the given work the 
fractal dimensionality is denoted in two ways: $D_{S}$ and $D_{div}$, where $D_{S}$ - 
fractal dimension value of whole studied picture. $D_{div}$ - means value of fractal 
dimension when the picture is divided in four parts. It should be noted (Fig.\ref{f6}), 
that $D_{div}$ < $D_{S}$, that in opinion of the authors, speaks about heterogeneity of 
structural fragments distributing in size. When fragment sizes approach some $<d>$ 
the size of $D_{div}$ will also tend to $D_{S}$. Also there is a deviation of points from 
the straight line because of the got structural pictures is not regular fractals 
\cite{vkb01}. That’s why values of fractal dimension vary the limit 
of variation makes $\pm0.03$. 

\begin{figure}
\hspace{0.06 cm}
\includegraphics [width=3 in]{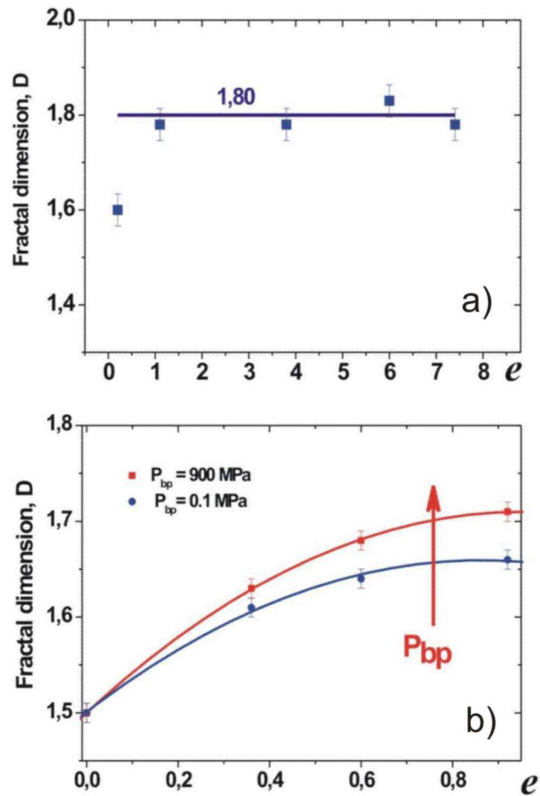}
\caption{\label{f7} Fractal dimension vs. plastic strain for iron wire (a) and molybdenum (b) cross section.
Also self-similarity effect (a) and pressure effect (b).
}
\end{figure}

Dependence of fractal dimension on plastic strain (cross section) for iron wire
 and molybdenum are presented in Fig.\ref{f7} a. During the deformation of iron wire there
 is the increase in fractal dimension of structural patterns (Fig.\ref{f6} a) 
from $1.60$ (full circle) at $e < 1 to 1.80$ (full squares) at $e>1$ and further 
maintenance at this level that speaks about self-similar structure 
evolution during severe plastic deformation. 

In the case of molybdenum (Fig.\ref{f7} b) 
the fractal dimension also increases from $1.60$ (without backpressure) to $1.70 $(with backpressure). Difference 
between the values of fractal dimension for molybdenum without backpressure 
and with backpressure, is because in the latter case there is a more intensive 
fragmentation of the material and more fine grain results from the same 
deformations than in the absence of backpressure. It follows that the finer 
the grain the higher fractal dimension in the case of a more uniform 
distribution of grains, that well correlates with theoretical data \cite{b05}. 
As a result, it can be concluded: 

\section{SHORT CONCLUSION}
In the initial undeformed or recrystallized metal, 
the same as in nonmetal model objects, the fractal
 dimension is permanent, $D=1.60\pm±0.03$. That speaks of 
the stochastic nature of structure evolution at the initial deformation stage; 

At growth of deformation the fractal dimension grows
 from $1.60\pm±0.03$ to $1.80\pm±0.03$ and remains at the given 
level with further growth of deformation thus speaking 
about the self-similar process of structure evolution; 

The fractal dimension increases with deformation pressure;

The fractal dimension of structures is an 
important quantitative characteristic of the 
deformed metal and undoubtedly requires further 
study. It is interesting to compare the fractal 
dimension with the exponent of degree in the Holl-Petch law.

\end{document}